\documentclass[10pt,twocolumn]{mc_nankai}

\usepackage[dvipsnames,table,xcdraw]{xcolor}
\usepackage{amsmath,amssymb,mathtools,bm}
\usepackage{booktabs}
\usepackage{colortbl}
\usepackage{enumitem}
\usepackage[numbers,sort&compress]{natbib}
\usepackage{xspace}

\definecolor{TableGroupBlue}{HTML}{D1E5F0}
\definecolor{QualitativeRowBlue}{HTML}{EAF4F8}
\definecolor{QualitativeGreen}{HTML}{238B45}
\definecolor{QualitativeRed}{HTML}{C43C39}
\newcommand{\qualgood}[1]{\textcolor{QualitativeGreen}{\textbf{#1}}}
\newcommand{\qualbad}[1]{\textcolor{QualitativeRed}{\textbf{#1}}}
\providecommand{\Description}[1]{}

\title{RecoReward: Recommender-Guided Multimodal Description Generation for Recommendation}
\author[1]{Guohong Mu}
\author[2]{Yueyang Liu}
\author[2,*]{Jiangxia Cao}
\author[2]{Changxin Lao}
\author[2]{Zijie Zhuang}
\author[2]{Yuhui Zhang}
\author[2]{Jiaqi Feng}
\author[3,4]{Ruochen Yang}
\author[2]{Shuang Yang}
\author[2]{Zhaojie Liu}
\author[1,*]{Qibin Hou}

\affiliation[1]{VCIP, School of Computer Science, Nankai University}
\affiliation[2]{Kuaishou Technology, Beijing, China}
\affiliation[3]{Institute of Information Engineering, Chinese Academy of Sciences, Beijing, China}
\affiliation[4]{School of Cyber Security, University of Chinese Academy of Sciences, Beijing, China}
\mclink[Contact]{\email{houqb@nankai.edu.cn, caojiangxia@kuaishou.com}}

\abstract{
Multimodal large language models (MLLMs) can convert multimodal item content into structured descriptions used as semantic features for recommendation.
Conventional content-only generation, however, cannot use downstream user signals to determine which semantics should be emphasized.
Recent user-conditioned methods incorporate these signals through user histories or profiles, but they require user information at inference and make generation user-dependent.
In this paper, we introduce RecoReward, which instead uses behavior-derived rewards during training and preserves content-only inference.
To instantiate this idea in live-stream recommendation, we treat historically engaged users as a proxy for future target users and use observational non-target users to estimate affinity shared broadly across users.
The Recommender Affinity Score (RAS) contrasts these signals to provide user-selective feedback for reinforcement learning, allowing the learned policy to generate a single shared description without user inputs.
In our offline benchmark, RecoReward-9B outperforms its Qwen3.5-9B baseline and all other evaluated models across seven recall metrics.
Online A/B testing also shows performance gains.
These results show that RecoReward trains the MLLM to produce item features that benefit downstream recommendation while retaining content-only serving.
}

\begin{document}
\maketitle
\begingroup
\renewcommand{\thefootnote}{*}
\footnotetext{Corresponding author.}
\endgroup
\justifying

\section{Introduction}
\label{sec:introduction}

Modern recommender systems~\citep{he2016vbpr,wei2019mmgcn,yu2023mgcn} increasingly use multimodal content to complement item identifiers, metadata, and historical interactions.
Images, video, speech, and text provide item semantics that conventional signals do not fully capture.
These semantics are important for content-aware matching, generalization, and cold-start recommendation.
MLLMs provide a flexible way to convert heterogeneous item content into structured descriptions in natural language.
Recommender systems can use these descriptions as item-side semantic features that complement conventional identifiers and pre-extracted content features~\citep{ren2024rlmrec,denadai2025describe,yang2026sarm}.
The resulting item interface is interpretable and compatible with existing two-tower architectures for candidate recall.

\begin{figure}[t]
  \centering
  \includegraphics[width=\columnwidth]{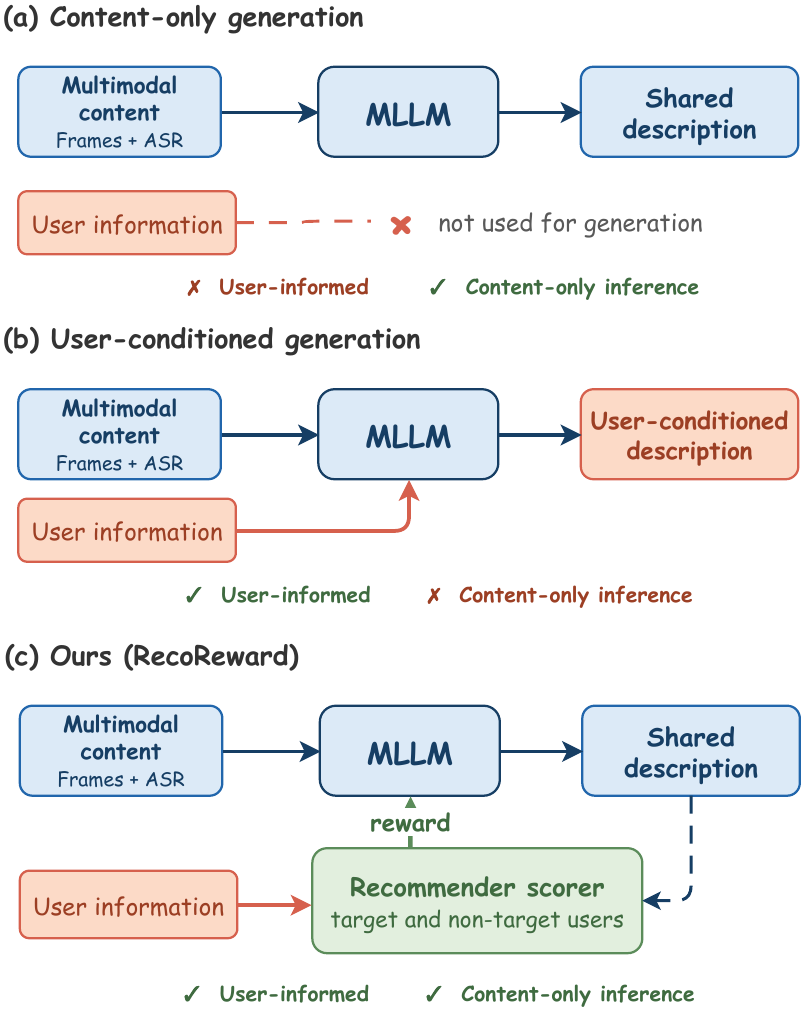}
  \caption{Content-only generation omits user-side guidance, whereas user-conditioned generation requires user information at inference.
  RecoReward introduces this information through the training reward and preserves content-only inference.}
  \label{fig:motivation}
\end{figure}

However, most content-oriented MLLM approaches generate descriptions from the observed item content alone and treat recommendation as a later consumer, rather than using recall-stage utility to guide what the description should emphasize~\citep{denadai2025describe,yang2026sarm}.
This omission matters in that a shared description is encoded once and reused when the item is compared with many users.
Content fidelity alone does not determine recommendation utility.
Descriptions that follow the same structured schema can still focus on different people, events, scenes, attributes, or user-relevant cues.
A fluent and complete description can therefore emphasize broadly relevant semantics while providing a weak signal for distinguishing the users most likely to engage with the item.

Recent recommender-guided methods show that user-side information can steer generated outputs toward downstream recommendation objectives.
Rec-R1~\cite{lin2025recr1} conditions generation on user queries or behavioral histories and optimizes the output with recommender feedback.
CURec~\cite{luo2025curec} uses interaction sequences to infer user interest patterns and generate personalized recommendation reasons, while VRAgent-R1~\cite{chen2025vragent} simulates user decisions from historical interactions.
These studies connect generation to user intent or behavior and report improvements in downstream recommendation.

However, making user information part of every MLLM request changes a reusable item representation into a user-dependent computation.
The serving system must access and serialize user histories or profiles, rebuild the prompt as interactions change, and repeat generation for different users instead of caching one item description.
Long and heterogeneous behavior sequences are also difficult for LLMs to interpret reliably.
ReLLa reports that performance can decline as histories grow even before the context limit is reached~\citep{lin2024rella}.
Figure~\ref{fig:motivation} summarizes this design trade-off.
Content-only generation is simple to serve but does not use user-side guidance, whereas direct user conditioning uses such guidance but requires user information during inference.
This comparison leads to our central question: \textbf{Can user supervision during training teach an MLLM to favor descriptions better suited to downstream recommendation while keeping inference content-only?}

\begin{figure}[t!]
  \centering
  \includegraphics[width=\columnwidth]{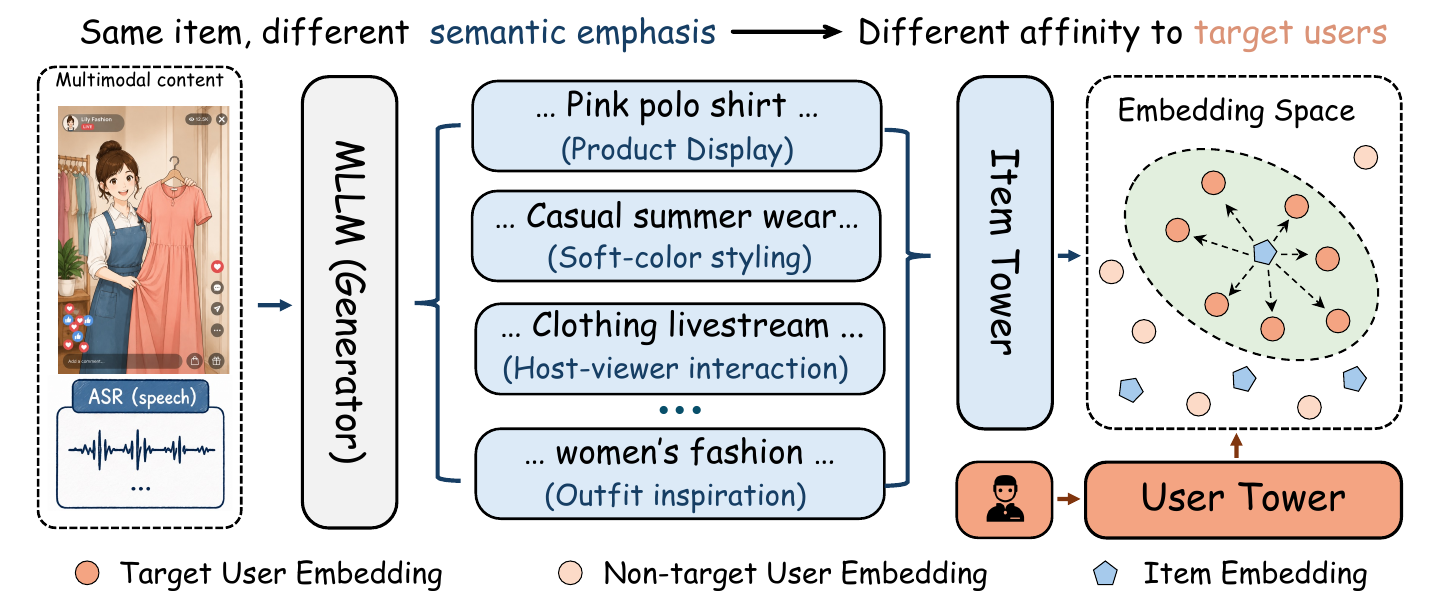}
  \caption{Key insight of RecoReward.
  Descriptions that fit the same multimodal item content can emphasize different semantics and produce different affinities for target users in the shared two-tower matching space.}
  \label{fig:insight}
\end{figure}

Our key insight is to use a behavior-trained DSSM as a training-time interface.
As Figure~\ref{fig:insight} shows, users and candidate descriptions share a matching space, allowing observed user behavior to score alternative descriptions without adding user information to the generator input.
RecoReward uses this interface to train a content-only generator, and we study it in live-stream recommendation.
Section~\ref{sec:empirical-motivation} validates the behavioral proxy and examines how non-target users help remove affinity shared broadly across users.

These findings motivate the Recommender Affinity Score (RAS), a user-selective recommender reward.
RAS contrasts compatibility with historically engaged target users against affinity shared with observational non-target users.
During training, user information remains confined to the frozen recommender-side scorer, and policy optimization internalizes its preference over candidate descriptions.
At serving time, the resulting policy receives only multimodal item content and generates one shared item-side description that can be reused across downstream recall decisions.
The formulation applies when a recommender can score users against generated item descriptions; our experiments evaluate this formulation in live-stream recommendation.

In offline evaluation for live-stream recommendation, both RecoReward policies improve all seven reported recall metrics over their corresponding base models.
RecoReward-9B achieves the highest mean among all evaluated models and improves the seven metrics by \(31.7\)--\(40.4\%\) relative to Qwen3.5-9B.
The reward-design results follow the predicted user-selectivity pattern, although they do not isolate this mechanism as the sole cause of policy improvement.
A system-level online A/B test further reports positive changes in key-page effective-user penetration and outflow distribution for RecoReward.

\noindent Our contributions are threefold:
\begin{list}{(\arabic{enumi})}{%
  \usecounter{enumi}
  \setlength{\leftmargin}{2em}
  \setlength{\labelwidth}{1.5em}
  \setlength{\labelsep}{0.5em}
  \setlength{\itemsep}{0pt}
  \setlength{\topsep}{4pt}
  \setlength{\parsep}{0pt}
  \setlength{\partopsep}{0pt}
  \renewcommand{\makelabel}[1]{#1\hfil}
}
  \item \textbf{Behavioral evidence for user-selective reward design.} In a temporally separated live-stream analysis, we show that historical target users predict future target users, while their mean representation also contains affinity shared with non-target users.
  \item \textbf{RAS-guided policy learning.} We formulate RAS to subtract this shared affinity and use it as sequence-level feedback for training a content-only MLLM policy. User features remain confined to the frozen recommender-side scorer.
  \item \textbf{Validation in live-stream recommendation.} We evaluate RecoReward through matched-tower recall evaluation and reward-design ablations. RecoReward-9B achieves the highest mean among all evaluated models on all seven recall metrics, improving by \(31.7\)--\(40.4\%\) over its base, and a separate system-level online A/B test reports positive changes in key-page and outflow metrics.
\end{list}

The remainder of the paper reviews related work, presents the behavioral analysis, formalizes RAS and policy optimization, and reports the offline and online evaluations.
\section{Related Work}
\label{sec:related-work}
\begingroup
\setlength{\emergencystretch}{2em}

\subsection{Multimodal Content Representation for Recommendation}

Multimodal recommendation uses visual, textual, acoustic, and video signals to complement discrete item identifiers, particularly in sparse-interaction and cold-start settings.
Early methods incorporated pretrained modality features into collaborative models, whereas subsequent approaches jointly modeled multimodal content and user-item structure or aligned general content representations with behavior-derived recommendation spaces~\citep{guo2024lgmrec,he2016vbpr,liu2024alignrec,wang2023missrec,wei2019mmgcn,wei2020grcn,yu2023mgcn,zhang2021lattice,zhou2023freedom,zhou2023bm3}.

Recent LLM- and MLLM-based methods formulate recommendation as language generation or connect semantic content with recommendation data, user behavior, and collaborative embeddings~\citep{bao2023tallrec,geng2022p5,kim2024allmrec,liao2024llara,lin2024rella}.
RLMRec aligns LLM-generated profiles with collaborative representations~\citep{ren2024rlmrec}, whereas NoteLLM-2 learns multimodal recommendation representations through vision-language adaptation~\citep{zhang2025notellm2}.
Other work encodes MLLM-generated video descriptions as recommender features~\citep{denadai2025describe}.
For live streaming, SARM generates semantic anchors and incorporates them into an industrial ranking model~\citep{yang2026sarm}.
Most of this work improves content representations or aligns generated semantics after generation.
RecoReward instead uses recommendation behavior to determine which content-compatible description the model generates as the shared item-side interface.

\subsection{Recommender-Guided Text Generation}

Several recent methods use recommender feedback to supervise language and multimodal generation.
Rec-R1 optimizes LLM outputs with a fixed black-box recommender for product search and sequential recommendation; CURec trains an LLM to generate personalized recommendation reasons and revises them with a recommendation-inspired reward model; and VRAgent-R1 combines multimodal item perception, user simulation, and reinforcement learning for video recommendation~\citep{lin2025recr1,luo2025curec,chen2025vragent}.
All three methods use recommendation objectives to guide generation, but their outputs and conditioning boundaries differ from those of RecoReward.
CURec produces user-conditioned reasons, VRAgent-R1 focuses on agent-based user simulation, and Rec-R1 provides a general closed-loop framework.
RecoReward generates a single shared item-side description from live content.
It derives supervision from author-level user sets, subtracts broadly shared affinity measured from observational non-target users, and requires no user information in the generator input at serving time.

\subsection{Reward-Guided Generation and Preference Optimization}

Sequence-level optimization addresses the mismatch between token-level likelihood and downstream utility.
Minimum Risk Training and Self-Critical Sequence Training optimize task-level generation metrics~\citep{rennie2017selfcritical,shen2016minimum}.
RLHF, DPO, and group-relative policy optimization extend reward-guided learning to open-ended language generation~\citep{ouyang2022instructgpt,rafailov2023dpo,shao2024deepseekmath,stiennon2020summarize}.

Optimizing a learned or approximate reward can eventually stop improving the intended objective~\citep{gao2023overoptimization}.
The reward in RecoReward differs from rewards for human preference, safety, text quality, and verifiable answers~\citep{kirstain2023pickapic,xu2023imagereward}.
It is derived from a recommender trained on observed behavior and measures the value of a description as an item-side feature for a historical-user proxy relative to a non-target population.
The resulting score is a recommender-space proxy, not a universal measure of description quality, factuality, or causal user preference.\par
\endgroup

\section{Behavioral Analysis for Reward Design}
\label{sec:empirical-motivation}

This section instantiates the general training framework in live-stream recommendation and examines why the historical target-user center is an incomplete proxy for recommendation-aligned description selection before policy optimization.
The analysis uses a frozen generator, a fixed candidate set, temporally separated user groups, and three frozen two-tower evaluators.

\subsection{Analysis Setup}
\label{sec:analysis-setup}

Each frozen two-tower evaluator comprises a user tower that encodes user histories and an item-side tower that encodes candidate descriptions.
The inner product of the two representations measures the compatibility between a user and a description.
Section~\ref{sec:reward-model} provides the full formulation and training objective.

For each author, we sample four disjoint groups: historical target users, historical non-target users, future target users, and future non-target users.
The historical target center defines the proxy direction, whereas subtracting the historical non-target center produces the corrected direction.
The future groups are reserved for evaluation.
The comparison holds the candidate set, user groups and tower parameters fixed.

\begin{figure*}[t!]
  \centering
  \includegraphics[width=\textwidth]{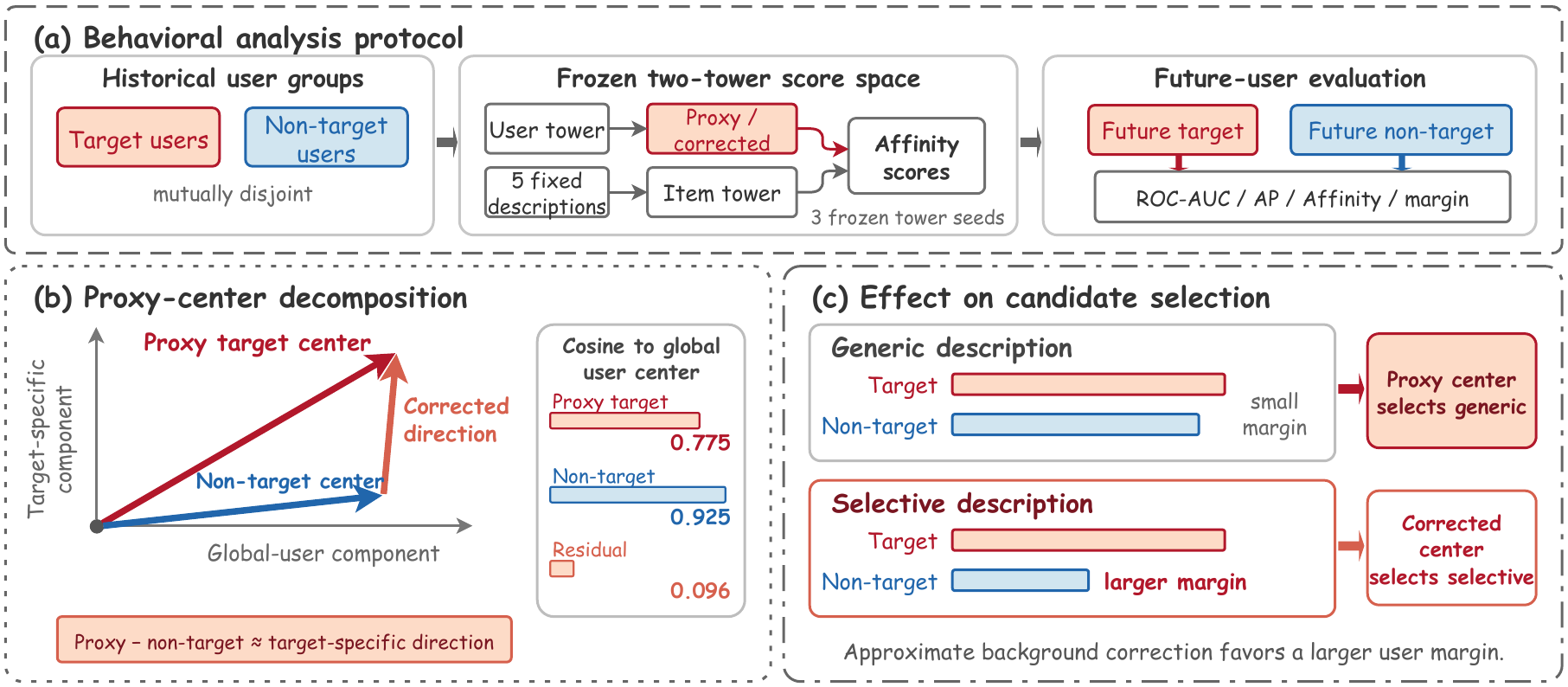}
  \caption{Behavioral analysis and proxy-center hypothesis.
  Panel (a) shows the temporally separated two-tower protocol.
  Panel (b) illustrates approximate removal of the global user component, and panel (c) shows how correction can favor a larger target--non-target margin.}
  \label{fig:section3-overview}
\end{figure*}

Figure~\ref{fig:section3-overview}(a) summarizes the temporally separated analysis protocol.

Table~\ref{tab:section3-metrics} defines four metric groups for candidate sensitivity, removal of shared affinity, temporal user prediction, and the target-user specificity of selected descriptions.

\subsection{Behavioral Findings}
\label{sec:behavioral-findings}

Figure~\ref{fig:section3-overview}(b) presents the proxy-center hypothesis and the global-alignment measurements.
The mean cosine similarity to the global user center is 0.7752 for the proxy target center and 0.9255 for the non-target center, indicating that both contain a strong global-user direction.
After subtraction, the residual direction has a cosine similarity of 0.0960 with the global center.
Together, these observations support the non-target center as a background estimate and are consistent with average cancellation of the shared component, but they do not establish an exact decomposition or a pure target center.

On the primary evaluation surface, the proxy target direction achieves a future-user ROC-AUC of 0.7603 and an AP of 0.7518.
The corrected direction increases these values to 0.8052 and 0.7904, respectively.
The proxy therefore retains future target-user information, while correction improves separation rather than simply removing the predictive signal.

Although all five candidates follow the same schema and describe the same stream, the proxy and corrected directions select different descriptions for 44.7--50.9\% of authors across the three evaluator seeds; the mean disagreement rate is 47.4\%.
Relative to proxy-center scoring, correction reduces future target affinity from 0.3108 to 0.2969, but it produces a larger reduction in future non-target affinity, from 0.1596 to 0.1354.
The future-user margin thus increases from 0.1513 to 0.1616.
The correction therefore changes candidate ordering and favors a more selective description, as illustrated in Figure~\ref{fig:section3-overview}(c); Table~\ref{tab:section3-summary} reports the exact values.

\subsection{Design Implications}
\label{sec:analysis-design-implications}

The historical target center is a useful but mixed proxy: it contains future target-user information together with affinity shared across broadly active users.
Scoring against this center alone can reward generic descriptions that match both target and non-target users.
The non-target center supplies a background term that reduces the shared affinity and yields a more target-specific proxy for candidate selection.

Historical target users provide proxies for future target users, not a complete or causal definition of user preference.
Under the sampling procedure, non-target users are active users without an observed positive interaction with the author.
The corrected direction is therefore an empirical proxy with reduced global alignment, not a true or fully purified target center.
Section~\ref{sec:ras} formalizes this correction at the affinity level.

\begin{table*}[t]
  \centering
  \begin{minipage}{\textwidth}
    \captionof{table}{Metrics used in the behavioral analysis and their corresponding interpretations.}
    \label{tab:section3-metrics}
    \small
    \setlength{\tabcolsep}{3pt}
    \begin{tabular*}{\linewidth}{@{\extracolsep{\fill}}p{0.18\linewidth}p{0.34\linewidth}p{0.39\linewidth}@{}}
      \toprule
      Metric & What it tests & Interpretation \\
      \midrule
      Disagreement rate
      & Whether subtraction changes the selected description
      & Higher means more frequent candidate changes \\
      Global-center cosine
      & Whether shared user affinity remains
      & Near zero means shared alignment is largely removed \\
      ROC-AUC / AP
      & Whether history separates future target and non-target users
      & Higher is better; ROC-AUC of 0.5 is random \\
      Future-user margin
      & Whether selection favors future target users
      & Larger values mean stronger target-user specificity \\
      \bottomrule
    \end{tabular*}
  \end{minipage}
\end{table*}

\begin{table}[t]
  \centering
  \setlength{\belowcaptionskip}{2pt}
  \caption{Future-user discrimination and selected-description affinity.}
  \label{tab:section3-summary}
  \scriptsize
  \setlength{\tabcolsep}{0pt}
  \renewcommand{\arraystretch}{0.90}
  \begin{tabular*}{\columnwidth}{@{\extracolsep{\fill}}lccccc@{}}
    \toprule
    & \multicolumn{2}{c}{Future-user discrimination}
    & \multicolumn{3}{c}{Selected-description affinity} \\
    \cmidrule(lr){2-3}\cmidrule(l){4-6}
    Direction & ROC-AUC & AP & Target & Non-target & Margin \\
    \midrule
    Proxy target & 0.7603 & 0.7518 & 0.3108 & 0.1596 & 0.1513 \\
    Corrected & 0.8052 & 0.7904 & 0.2969 & 0.1354 & 0.1616 \\
    \bottomrule
  \end{tabular*}
\end{table}

\section{Method}
\label{sec:method}

\begin{figure*}[t]
  \centering
  \includegraphics[width=\textwidth]{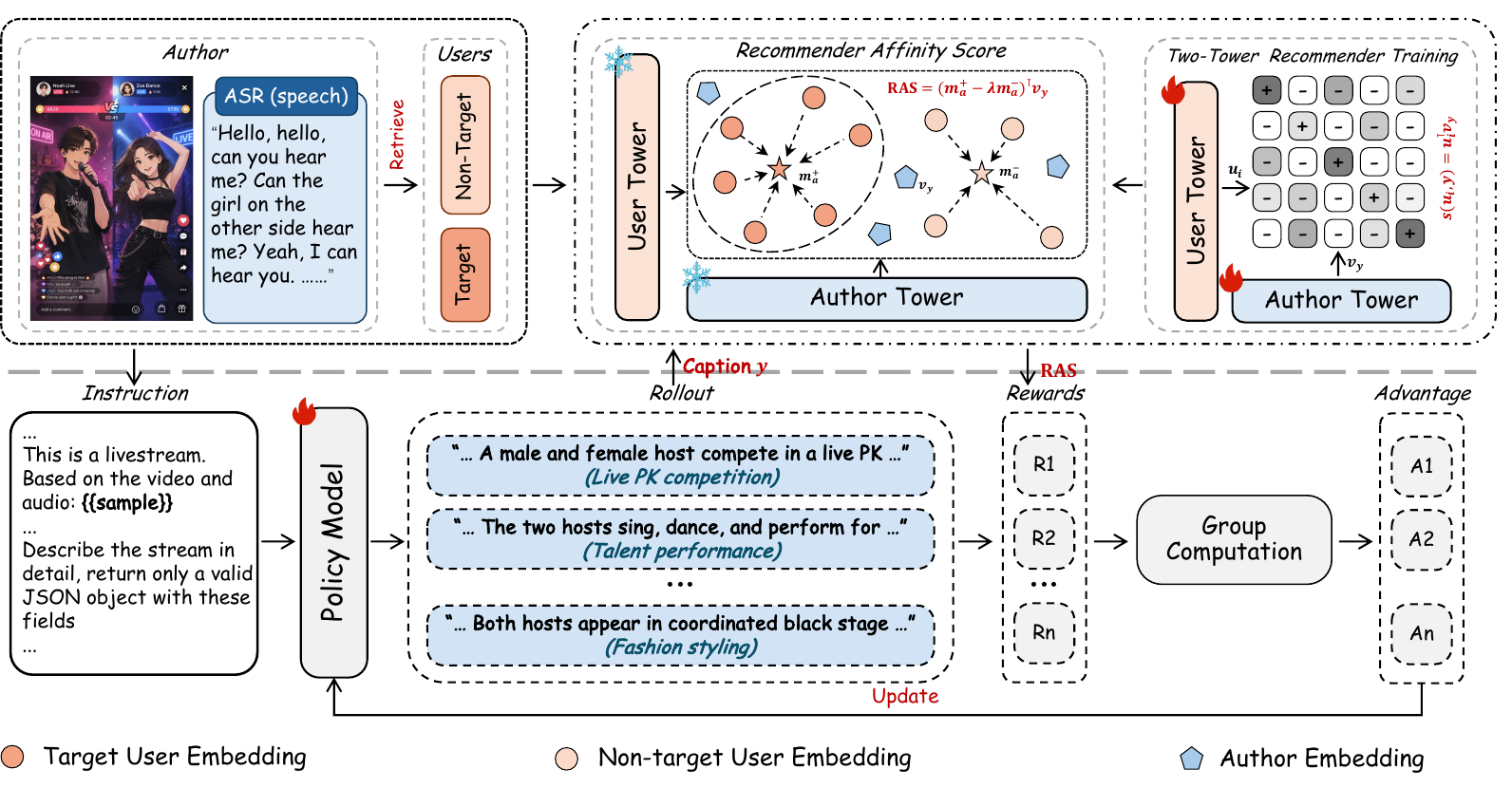}
  \caption{Training pipeline of RecoReward.
  The policy samples candidate descriptions from live-stream content.
  The frozen two-tower recommender assigns RAS rewards, and group-relative optimization updates the policy.}
  \Description{Training pipeline of RecoReward with live-stream content as policy input, multiple generated descriptions, RAS rewards from a frozen two-tower scorer, and group-relative optimization of the policy.}
  \label{fig:method-overview}
\end{figure*}

The empirical analysis identifies a predictive historical-user signal and a shared-affinity component that should be removed.
This section formalizes the contrastive construction as the Recommender Affinity Score (RAS) and applies the resulting scalar feedback to generator optimization.
Figure~\ref{fig:method-overview} presents the complete training pipeline: content-only generation, recommender scoring, and group-relative policy updates.
Behavioral user features remain confined to the frozen recommender-side scorer and never enter the MLLM input.

\FloatBarrier

\subsection{Problem Formulation and Preliminaries}
\label{sec:problem-formulation}

Let \(x=(F,T)\) denote a live-stream context consisting of visual frames \(F\) and transcribed speech segments \(T\).
The MLLM policy generates a structured description according to
\begin{equation}
y\sim\pi_\theta(\cdot\mid x).
\label{eq:generation-policy}
\end{equation}
Let \(\mathcal{Y}(x)\) be the set of descriptions that are supported by the observed content and conform to the required output schema.
For author \(a\), the ideal description solves
\begin{equation}
y^*
\in
\arg\max_{y\in\mathcal{Y}(x)}
U_{\mathrm{rec}}(y;a),
\label{eq:recommendation-objective}
\end{equation}
where \(U_{\mathrm{rec}}\) denotes the recommendation utility obtained by using \(y\) as a shared item-side semantic feature.
The optimization is restricted to the content and format constraints represented by \(\mathcal{Y}(x)\).
Because the true utility is unavailable during policy training, RAS is a behavior-derived proxy rather than an equivalent definition of \(U_{\mathrm{rec}}\).

Reinforcement learning is a common approach to learning from preference signals.
GRPO samples \(G\) responses for each input and normalizes the rewards within the group, which enables relative optimization without a learned critic.
DAPO adds an asymmetrically clipped token-level objective and averages the loss over valid response tokens~\citep{shao2024deepseekmath,yu2025dapo}.
These methods provide the optimization background for the RAS-guided training introduced below.

\subsection{Behavior-Aligned Two-Tower Space}
\label{sec:reward-model}

The frozen scorer comprises a user tower and a description tower that map behavioral context and generated text into a shared recommendation space.
For user \(i\), let \(h_i\) denote the complete user-side feature record, including the identifier, attributes, and recent behavior sequence of the user.
For a generated description \(y\), the description tower receives only the corresponding text tokens.
Both towers produce normalized representations in the same \(d\)-dimensional space:
\begin{equation}
u_i=f_u(h_i),
\qquad
v_y=f_i(y),
\qquad
s(u_i,y)=u_i^\top v_y.
\label{eq:tower-score}
\end{equation}
The inner product \(s(u_i,y)\) measures the compatibility between user \(i\) and description \(y\) learned from observed behavior.
Author identifiers and author-specific ID embeddings are excluded from \(v_y\), which prevents identity features from directly determining score differences among descriptions of the same stream.

The towers are trained on positive user-stream interactions.
Consider a mini-batch \(\mathcal{B}=\{(h_b,y_b,w_b)\}_{b=1}^{N_{\mathrm{mb}}}\), where \(N_{\mathrm{mb}}=|\mathcal{B}|\) denotes the mini-batch size.
User \(b\) is paired with the description of an interacted stream, whereas the remaining batch descriptions are contrastive items.
With \(u_b=f_u(h_b)\) and \(v_j=f_i(y_j)\), the matching probability is
\begin{equation}
p(y_b\mid h_b;\mathcal{B})
=
\frac{\exp(u_b^\top v_b/\tau)}
{\sum_{j=1}^{N_{\mathrm{mb}}}\exp(u_b^\top v_j/\tau)}.
\label{eq:tower-probability}
\end{equation}
Training minimizes the behavior-weighted objective
\begin{equation}
\mathcal{L}_{\mathrm{tower}}
=
-\sum_{b=1}^{N_{\mathrm{mb}}}w_b\log p(y_b\mid h_b;\mathcal{B}).
\label{eq:tower-loss}
\end{equation}
The weight \(w_b\) represents the strength of observed positive behaviors, including room entry, long viewing, following, commenting, liking, and gifting.
The objective shapes the matching space according to recommendation behavior rather than generic language similarity.
\subsection{Recommender Affinity Score}
\label{sec:ras}

RAS converts the frozen user-description matching function into a user-selective scalar reward.
It rewards compatibility with historically engaged target users while subtracting affinity shared with non-target users.

\subsubsection{Target and Non-Target User Sets}

For author \(a\), the target set comprises users who recently exhibited a specified positive behavior toward the author, whereas the non-target set comprises active users with no observed positive interaction with the author during the construction window:
\begin{equation}
\mathcal{U}_a^+
=
\{u_{a,i}^+\}_{i=1}^{M},
\qquad
\mathcal{U}_a^-
=
\{u_{a,j}^-\}_{j=1}^{B}.
\label{eq:user-sets}
\end{equation}
The target set is an empirical historical proxy for the potential future users of author \(a\), not a complete or causal definition of user preference.
Non-target users form an observational population; they are not labeled as exposed negatives, users who expressed dislike, or counterfactual user labels.
The two-tower scorer and materialized user representations remain fixed throughout each policy-training run.

\subsubsection{User-Selective Compatibility}

Let the empirical target and non-target centers be
\begin{equation}
m_a^+
=
\frac{1}{M}\sum_{i=1}^{M}u_{a,i}^+,
\qquad
m_a^-
=
\frac{1}{B}\sum_{j=1}^{B}u_{a,j}^-.
\label{eq:user-centers}
\end{equation}
For description \(y\), the corresponding affinity components are
\begin{equation}
S^+(y;a)
=
\frac{1}{M}\sum_{i=1}^{M}s(u_{a,i}^+,y)
=
(m_a^+)^\top v_y,
\label{eq:target-affinity}
\end{equation}
\begin{equation}
S^-(y;a)
=
\frac{1}{B}\sum_{j=1}^{B}s(u_{a,j}^-,y)
=
(m_a^-)^\top v_y.
\label{eq:nontarget-affinity}
\end{equation}
These quantities are intermediate components, not independent method objectives.

The Recommender Affinity Score is defined as
\begin{equation}
\begin{aligned}
\operatorname{RAS}_{\lambda}(y;a)
&= S^+(y;a)-\lambda S^-(y;a) \\
&= (m_a^+-\lambda m_a^-)^\top v_y,
\end{aligned}
\label{eq:ras}
\end{equation}
where \(\lambda\geq0\) controls the strength of non-target subtraction.
The target and non-target terms implement the matching and shared-affinity subtraction requirements from Section~\ref{sec:analysis-design-implications}, respectively.
RAS thus measures user-selective compatibility in the behavior-trained recommendation space.
It does not measure factuality, estimate causal preference, or represent universal recommendation utility.
\subsection{Reinforcement Learning}
\label{sec:reinforcement-learning}

RAS assigns a sequence-level reward to each content-compatible description.
Group-relative updates use these rewards to learn which semantic choices benefit recommendation.
The policy stores the resulting ordering in its parameters and generates one description from content alone at serving time.

\subsubsection{Reward Construction}

Because the normalized tower affinities lie in \([-1,1]\), \(\operatorname{RAS}_{\lambda}\) has the theoretical range \([-(1+\lambda),1+\lambda]\).
We map this range to the unit interval:
\begin{equation}
r_{\mathrm{sem}}(y;a)
=
\operatorname{clip}_{[0,1]}
\left(
\frac{\operatorname{RAS}_{\lambda}(y;a)+(1+\lambda)}
{2(1+\lambda)}
\right).
\label{eq:ras-reward}
\end{equation}
Let \(r_{\mathrm{fmt}}(y)\in\{0,1\}\) indicate whether the output is a valid JSON object that contains every required field.
The final reward combines the two terms:
\begin{equation}
r(y;a)
=
\alpha r_{\mathrm{sem}}(y;a)
+(1-\alpha)r_{\mathrm{fmt}}(y).
\label{eq:final-reward}
\end{equation}
The semantic term measures user-selective recommender compatibility, whereas the format term enforces syntactic parseability without establishing factual grounding.

\subsubsection{Training and Serving Boundary}
\label{sec:information-boundary}

Only the frozen scorer accesses historical target and non-target users during training.
The MLLM receives live-stream frames, transcribed speech, and the scalar reward induced by the output; it never receives user identifiers, histories, or embeddings as input.
After optimization, the policy generates a single shared description directly from live-stream content without accessing user information or invoking the scorer.
The learned model is a content-only policy trained with behavior-derived user supervision, not a personalized generator for individual users.

\begin{table*}[t!]
  \centering
  \caption{Overall recall performance on our benchmark.
  Entries (\%) are reported as the mean \(\pm\) population standard deviation across matched evaluators.
  Bold values indicate the highest mean in each column.}
  \label{tab:main-results}
  \scriptsize
  \setlength{\tabcolsep}{1.5pt}
  \renewcommand{\arraystretch}{1.05}
  \resizebox{\textwidth}{!}{%
  \begin{tabular}{@{}lccccccc@{}}
    \toprule
    Model
    & HR@10~\(\uparrow\)
    & HR@64~\(\uparrow\)
    & HR@128~\(\uparrow\)
    & NDCG@10~\(\uparrow\)
    & NDCG@64~\(\uparrow\)
    & NDCG@128~\(\uparrow\)
    & MRR~\(\uparrow\) \\
    \midrule
    \rowcolor{TableGroupBlue}
    \multicolumn{8}{@{}l}{\textbf{Non-MLLM Methods}} \\
    DSSM~\citep{huang2013dssm} & \(0.2292\pm0.0099\) & \(1.3858\pm0.0207\) & \(2.6884\pm0.0722\) & \(0.1042\pm0.0036\) & \(0.3336\pm0.0019\) & \(0.5322\pm0.0128\) & \(0.1725\pm0.0025\) \\
    LightGCN~\citep{he2020lightgcn} & \(0.3009\pm0.0188\) & \(1.5724\pm0.0389\) & \(2.9404\pm0.0451\) & \(0.1606\pm0.0140\) & \(0.4149\pm0.0053\) & \(0.6234\pm0.0035\) & \(0.2296\pm0.0102\) \\
    NextItNet~\citep{yuan2019nextitnet} & \(0.5865\pm0.1091\) & \(3.2467\pm0.1196\) & \(6.2128\pm0.5687\) & \(0.2888\pm0.0722\) & \(0.8191\pm0.0374\) & \(1.2717\pm0.0347\) & \(0.4149\pm0.0540\) \\
    GRU4Rec~\citep{hidasi2016gru4rec} & \(0.5791\pm0.1254\) & \(3.0671\pm0.0893\) & \(5.9253\pm0.2929\) & \(0.2785\pm0.0823\) & \(0.7781\pm0.0787\) & \(1.2138\pm0.0692\) & \(0.3989\pm0.0660\) \\
    \midrule
    \rowcolor{TableGroupBlue}
    \multicolumn{8}{@{}l}{\textbf{Commercial Models + DSSM}} \\
    GPT-5~\citep{openai2025gpt5} & \(0.7790\pm0.0367\) & \(3.7555\pm0.1743\) & \(6.4385\pm0.2625\) & \(0.3445\pm0.0180\) & \(0.9477\pm0.0462\) & \(1.3577\pm0.0597\) & \(0.4307\pm0.0189\) \\
    Gemini-3.1-Pro~\citep{google2026gemini31} & \(0.8943\pm0.0247\) & \(4.0332\pm0.0777\) & \(6.9300\pm0.1660\) & \(0.4019\pm0.0132\) & \(1.0371\pm0.0201\) & \(1.4793\pm0.0310\) & \(0.4841\pm0.0117\) \\
    \midrule
    \rowcolor{TableGroupBlue}
    \multicolumn{8}{@{}l}{\textbf{Open-Source Models + DSSM}} \\
    Qwen3-VL-4B-Instruct~\citep{bai2025qwen3vl} & \(0.7996\pm0.0245\) & \(4.0215\pm0.0393\) & \(7.1210\pm0.0683\) & \(0.3561\pm0.0042\) & \(1.0081\pm0.0088\) & \(1.4814\pm0.0134\) & \(0.4596\pm0.0011\) \\
    Qwen3-VL-8B-Instruct~\citep{bai2025qwen3vl} & \(0.8181\pm0.0415\) & \(4.1787\pm0.1350\) & \(7.1849\pm0.1888\) & \(0.3664\pm0.0245\) & \(1.0444\pm0.0435\) & \(1.5028\pm0.0470\) & \(0.4677\pm0.0219\) \\
    Qwen3.5-4B~\citep{qwen2026qwen35} & \(0.7721\pm0.0927\) & \(3.6704\pm0.3507\) & \(6.4266\pm0.5331\) & \(0.3517\pm0.0401\) & \(0.9371\pm0.0928\) & \(1.3577\pm0.1205\) & \(0.4424\pm0.0367\) \\
    Qwen3.5-9B~\citep{qwen2026qwen35} & \(0.7577\pm0.0238\) & \(3.6026\pm0.1168\) & \(6.2489\pm0.1832\) & \(0.3392\pm0.0076\) & \(0.9129\pm0.0265\) & \(1.3166\pm0.0359\) & \(0.4259\pm0.0079\) \\
    InternVL3-8B~\citep{zhu2025internvl3} & \(0.6894\pm0.0688\) & \(3.3061\pm0.2219\) & \(5.8540\pm0.2906\) & \(0.3118\pm0.0305\) & \(0.8401\pm0.0641\) & \(1.2284\pm0.0745\) & \(0.4004\pm0.0270\) \\
    InternVL3.5-4B~\citep{wang2025internvl35} & \(0.6414\pm0.0663\) & \(3.3502\pm0.2930\) & \(5.9787\pm0.4395\) & \(0.2780\pm0.0276\) & \(0.8244\pm0.0731\) & \(1.2254\pm0.0953\) & \(0.3750\pm0.0254\) \\
    InternVL3.5-8B~\citep{wang2025internvl35} & \(0.6731\pm0.0701\) & \(3.5165\pm0.2133\) & \(6.2287\pm0.3731\) & \(0.2973\pm0.0332\) & \(0.8681\pm0.0619\) & \(1.2816\pm0.0842\) & \(0.3971\pm0.0303\) \\
    \midrule
    \rowcolor{TableGroupBlue}
    \multicolumn{8}{@{}l}{\textbf{Ours + DSSM}} \\
    RecoReward-4B & \(0.9211\pm0.0864\) & \(4.8315\pm0.2684\) & \(8.4187\pm0.4224\) & \(0.4177\pm0.0461\) & \(1.2095\pm0.0844\) & \(1.7572\pm0.1074\) & \(0.5391\pm0.0436\) \\
    RecoReward-9B & \(\mathbf{0.9978\pm0.0289}\) & \(\mathbf{5.0577\pm0.1379}\) & \(\mathbf{8.6597\pm0.1871}\) & \(\mathbf{0.4509\pm0.0257}\) & \(\mathbf{1.2701\pm0.0416}\) & \(\mathbf{1.8199\pm0.0483}\) & \(\mathbf{0.5646\pm0.0271}\) \\
    \bottomrule
  \end{tabular}
  }
\end{table*}

\section{Experiments}
\label{sec:experiments}

In this section, we conduct offline and online experiments to answer the following research questions:
\begin{list}{\textbullet}{%
  \setlength{\leftmargin}{1.8em}
  \setlength{\labelwidth}{1.3em}
  \setlength{\labelsep}{0.5em}
  \setlength{\itemsep}{0pt}
  \setlength{\topsep}{2pt}
  \setlength{\parsep}{0pt}
  \setlength{\partopsep}{0pt}
  \renewcommand{\makelabel}[1]{#1\hfil}
}
  \item \textbf{RQ1:} How does RecoReward compare with representative MLLMs in downstream recall?
  \item \textbf{RQ2:} How does the downstream recall performance of RecoReward change under different hyperparameter settings?
  \item \textbf{RQ3:} How does RecoReward affect an online live-streaming recommendation system?
\end{list}

\subsection{Experimental Settings}

\subsubsection{Dataset}

The offline dataset is constructed from one week of observed user behavior on the Kuaishou live-streaming platform.
We use the first six days for training and hold out the final day as the test set for offline recall evaluation, ensuring a strictly chronological split.
The training data contain 69,947 live-stream items, 7,392,141 positive interactions, 1,042,130 users, and 56,443 authors.
For each query, the positive item is a live stream not observed during training and is ranked against a shared candidate universe.
Other known positive interactions of the same user are masked so that the evaluation does not treat them as negatives.

\subsubsection{Policy Training}

We train 4B and 9B policies initialized from the corresponding Qwen3.5 models.
The vision encoder is frozen, whereas all language-model parameters are updated with Fully Sharded Data Parallel training.
Policy training uses 8,000 live-stream items selected for complete multimodal inputs and sufficient user feedback.
None of these RL training items appears in the evaluation set.
Under the fixed sampling procedure, non-target users are active users with no observed positive interaction with the current author.
Each configuration processes one fixed-order epoch for 500 optimizer steps.
Optimization uses a learning rate of \(1\times10^{-6}\), an asymmetric clipping interval of \([1-0.2,1+0.28]\), a rollout temperature of 1, top-\(p\) of 1, and no top-\(k\) truncation.
The semantic and format reward weights are 0.9 and 0.1, respectively.
Training uses eight GPUs, each with 80 GB of memory.

\subsubsection{Evaluation}

For each generator, we regenerate descriptions for the same 69,947 live-stream items and train new DSSMs on the resulting description distribution.
For each generator, we train three DSSMs for 3 epochs with a different random seed, a global batch size of 8,192, a learning rate of \(1\times10^{-3}\), a weight decay of \(1\times10^{-4}\), and a contrastive temperature of 0.07.
The random seeds measure dispersion due to evaluator initialization rather than independent actor training.
All models share the interactions, architecture, optimization settings, and frozen next-day recall benchmark.

We report HR@K and NDCG@K for \(K\in\{10,64,128\}\), together with MRR; Appendix~\ref{app:experimental-details} provides the complete definitions.
Values are the mean and population standard deviation across evaluator seeds.

\vspace{-10pt}
\subsection{RQ1: Overall Recommendation Performance}
\label{sec:rq1}

We compare RecoReward with representative non-MLLM recommendation methods and MLLMs under a common matched-tower evaluation protocol.

RecoReward-9B leads all seven metrics, outperforming every non-MLLM baseline.
Relative to Qwen3.5-9B, it raises NDCG@128 from 0.013166 to 0.018199 and HR@128 from 0.062489 to 0.086597, with gains of \(31.7\)--\(40.4\%\) across all seven metrics.
RecoReward-4B also remains stronger than every evaluated unoptimized 8B and 9B model.
Moreover, among the open-source baselines, recall performance does not increase monotonically with model size: Qwen3.5-9B trails Qwen3.5-4B on all seven metrics.
This result shows that larger model scale does not guarantee higher downstream recommendation utility when downstream behavior is absent from the generation objective.
It therefore supports our motivation for incorporating downstream recommendation behavior into the training objective.
Appendix~\ref{app:qualitative} provides four qualitative comparisons that show how the five models place different emphasis on events and target users for the same multimodal input.

\subsection{RQ2: Hyperparameter Ablation}
\label{sec:rq2}

RQ2 examines three settings that affect reward construction and policy optimization: the non-target subtraction coefficient, the rollout count, and the number of users included in reward computation.
\subsubsection{Non-Target Subtraction Coefficient}
\label{sec:rq2-lambda}

We vary the non-target subtraction coefficient while holding the data, model initialization, and evaluation protocol fixed within the ablation.
The target-only configuration provides the no-subtraction baseline, whereas positive coefficients introduce non-target subtraction.
For each configuration, Table~\ref{tab:lambda-ablation} reports the checkpoint retained under the fixed selection protocol.

\begin{table}[t]
  \centering
  \caption{Recall performance for different non-target subtraction coefficients \(\lambda\).
  Entries (\%) are overall mean scores across three evaluator
seeds.}
  \label{tab:lambda-ablation}
  \scriptsize
  \setlength{\tabcolsep}{1.5pt}
  \renewcommand{\arraystretch}{1.12}
  \resizebox{\columnwidth}{!}{%
  \begin{tabular}{@{}cccccc@{}}
    \toprule
    \(\lambda\)
    & NDCG@64\(\uparrow\)
    & NDCG@128\(\uparrow\)
    & HR@64\(\uparrow\)
    & HR@128\(\uparrow\)
    & MRR\(\uparrow\) \\
    \midrule
    0 & 0.9663 & 1.4118 & 3.8698 & 6.7874 & 0.4417 \\
    0.5 & 1.0857 & 1.5716 & 4.3333 & 7.5160 & 0.4856 \\
    1 & 1.0755 & 1.5291 & 4.2296 & 7.2027 & 0.4906 \\
    2 & \textbf{1.2095} & \textbf{1.7572} & \textbf{4.8315} & \textbf{8.4187} & \textbf{0.5391} \\
    \bottomrule
  \end{tabular}
  }
\end{table}

All policies with positive coefficients outperform the target-only row on the five displayed metrics, and \(\lambda=2\) yields the strongest retained policy across all measured recall metrics.
Relative to \(\lambda=0\), this configuration increases NDCG@128 from 0.014118 to 0.017572, HR@128 from 0.067874 to 0.084187, and MRR from 0.004417 to 0.005391.
Relative to \(\lambda=1\), \(\lambda=0.5\) applies a milder non-target penalty and retains more target-affinity signal, which may explain its higher NDCG@64, NDCG@128, HR@64, and HR@128.
The higher MRR at \(\lambda=1\) suggests that stronger subtraction sharpens the top-ranked ordering despite lower performance at deeper cutoffs.
At \(\lambda=2\), aggressive suppression of descriptions that match both user groups produces a more user-selective reward ordering and clearer relative advantages for policy learning, consistent with Section~\ref{sec:behavioral-findings}.
The non-monotonic comparison between \(\lambda=0.5\) and \(\lambda=1\) shows that increasing \(\lambda\) does not always improve performance.

\subsubsection{Rollout Count}
\label{sec:rq2-rollouts}

The rollout count \(G\) determines both the number of descriptions sampled for each input and the number of samples used to estimate group-relative advantages.
We compare \(G\in\{4,8,12,16\}\) with fixed batch size, data order, learning rate, optimizer-step budget, and evaluation protocol.

\begin{table}[t]
  \centering
  \caption{Recall performance for different rollout counts \(G\).
  Entries (\%) are overall mean scores across three evaluator
seeds.}
  \label{tab:rollout-ablation}
  \scriptsize
  \setlength{\tabcolsep}{1.5pt}
  \renewcommand{\arraystretch}{1.12}
  \resizebox{\columnwidth}{!}{%
  \begin{tabular}{@{}cccccc@{}}
    \toprule
    \(G\)
    & NDCG@64\(\uparrow\)
    & NDCG@128\(\uparrow\)
    & HR@64\(\uparrow\)
    & HR@128\(\uparrow\)
    & MRR\(\uparrow\) \\
    \midrule
    4 & 1.0575 & 1.5099 & 4.1435 & 7.1072 & 0.4872 \\
    8 & 1.0755 & 1.5291 & 4.2296 & 7.2027 & 0.4906 \\
    12 & \textbf{1.1770} & 1.6528 & \textbf{4.6122} & 7.7258 & \textbf{0.5315} \\
    16 & 1.1516 & \textbf{1.6606} & 4.5647 & \textbf{7.8990} & 0.5182 \\
    \bottomrule
  \end{tabular}
  }
\end{table}

Every rollout setting in Table~\ref{tab:rollout-ablation} outperforms the Qwen3.5-4B baseline in Table~\ref{tab:main-results} on the five displayed metrics.
Performance improves on every displayed metric as \(G\) increases from 4 to 12.
A plausible explanation is that a larger rollout group covers a broader set of descriptions and provides richer within-group comparisons, increasing the chance of observing useful high-reward outputs and producing more informative relative advantages.
The trend does not continue uniformly at \(G=16\): NDCG@128 and HR@128 improve, whereas NDCG@64, HR@64, and MRR decline.
With more samples, the group is also more likely to contain outputs from the high-RAS tail.
These outputs can receive strong positive advantages and push the policy toward patterns favored by the proxy reward.
Because RAS approximates downstream recommendation utility rather than measuring it directly, stronger optimization of such tail behavior may reduce generalization.
The results suggest that \(G=12\) provides a better balance between output exploration and over-optimization of the reward proxy under the current setting.
This interpretation remains descriptive because each configuration uses one actor run and larger \(G\) also increases the total number of sampled trajectories.

\subsubsection{Reward User Count}
\label{sec:rq2-reward-users}

The user cap \(M\) controls the maximum number of target users and the equal number of non-target users averaged on each side of RAS.
We compare \(M\in\{25,50,100,200\}\) while holding the live-stream items, user ordering, model initialization, rollout count, optimizer settings, and matched-tower evaluation protocol fixed.

\begin{table}[t]
  \centering
  \caption{Recall performance for different reward user caps \(M\).
  Entries (\%) are overall mean scores across three evaluator seeds.}
  \label{tab:reward-user-ablation}
  \scriptsize
  \setlength{\tabcolsep}{1.5pt}
  \renewcommand{\arraystretch}{1.12}
  \resizebox{\columnwidth}{!}{%
  \begin{tabular}{@{}cccccc@{}}
    \toprule
    \(M\)
    & NDCG@64\(\uparrow\)
    & NDCG@128\(\uparrow\)
    & HR@64\(\uparrow\)
    & HR@128\(\uparrow\)
    & MRR\(\uparrow\) \\
    \midrule
    25 & \textbf{1.1636} & \textbf{1.6475} & \textbf{4.5281} & \textbf{7.6966} & \textbf{0.5376} \\
    50 & 1.1230 & 1.6087 & 4.4232 & 7.6085 & 0.5110 \\
    100 & 1.0755 & 1.5291 & 4.2296 & 7.2027 & 0.4906 \\
    200 & 1.0606 & 1.5015 & 4.1670 & 7.0564 & 0.4842 \\
    \bottomrule
  \end{tabular}
  }
\end{table}

The \(M=25\) configuration is strongest on all five displayed metrics, followed by \(M=50\), \(M=100\), and \(M=200\).
Relative to \(M=100\), \(M=25\) increases NDCG@128 from 0.015291 to 0.016475, HR@128 from 0.072027 to 0.076966, and MRR from 0.004906 to 0.005376.
The results suggest a trade-off between user selectivity and estimation stability.
Although increasing \(M\) produces more stable affinity estimates, averaging over a broader user set may weaken the target--non-target contrast and make the reward less informative for policy learning.
Within the tested range, \(M=25\) appears to preserve a sharper user-selective signal.

\subsection{RQ3: Online A/B Test}
\label{sec:rq3}

\begin{table}[t]
  \centering
  \caption{Results of the one-week online A/B test for RecoReward.}
  \label{tab:online-results}
  \footnotesize
  \setlength{\tabcolsep}{0pt}
  \renewcommand{\arraystretch}{1.05}
  \begin{tabular*}{\columnwidth}{@{\extracolsep{\fill}}lccc@{}}
    \toprule
    & \multicolumn{1}{c}{Key Pages}
    & \multicolumn{2}{c}{Outflow} \\
    \cmidrule(lr){2-2}
    \cmidrule(lr){3-4}
    System
    & Effective-User Penetration
    & Exposure
    & Users \\
    \midrule
    Kuaishou
    & \(+0.265\%\)
    & \(+0.791\%\)
    & \(+0.740\%\) \\
    \bottomrule
  \end{tabular*}
\end{table}

We evaluate RecoReward through a one-week, system-level A/B test on the Kuaishou live-streaming service.
Effective-user penetration measures the share of users who meet the platform-defined effective-viewing criterion on key pages.
Outflow exposure and users measure the number of impressions and unique users reached through outflow traffic, respectively.
Table~\ref{tab:online-results} reports the changes in these metrics.
RecoReward increases effective-user penetration on key pages by \(0.265\%\), while improving outflow exposure and users by \(0.791\%\) and \(0.740\%\), respectively.
These results show that recommendation-aligned semantic modeling improves user engagement on key pages and expands live-stream distribution in real-world deployment.

\section{Conclusion}
\label{sec:conclusion}

We present RecoReward, a training framework that uses recommender feedback to improve multimodal item descriptions while retaining content-only inference.
A frozen two-tower scorer evaluates candidates against historical target and non-target users, and RAS reduces broadly shared affinity to provide a user-selective reward for policy learning.

RecoReward-9B ranks first across all seven recall metrics, improving by \(31.7\)--\(40.4\%\) over Qwen3.5-9B.
The ablations support non-target subtraction and reveal trade-offs in the rollout count and reward user count.
In a one-week online A/B test, RecoReward improves key-page effective-user penetration and expands outflow reach.
Together, these results show that downstream recommendation behavior can train an MLLM to generate more useful item-side features without user-dependent serving, although the evidence remains limited to observational user groups and live-stream recommendation.

\bibliographystyle{ieee_fullname}
\bibliography{references}

\clearpage
\appendix
\section{Full Generation Prompt}
\label{app:prompt}

Table~\ref{tab:full-generation-prompt} presents an English translation of the complete Chinese prompt used for the reported base-model and policy evaluations.
The bracketed placeholders indicate where the ordered screenshots and ASR segments are inserted.
User identities, histories, and representations are not included in the generator prompt.
The seven JSON field names are translated for readability; the implementation used the corresponding Chinese keys.

\begin{table*}[!t]
\centering
\caption{English translation of the complete generation prompt used in the experiments.}
\label{tab:full-generation-prompt}
\setlength{\fboxsep}{6pt}
\noindent\fcolorbox{black!45}{black!2}{%
\begin{minipage}{0.95\textwidth}
\begingroup
\small
\raggedright
\setlength{\parindent}{0pt}
\setlength{\parskip}{2pt}
This is a live stream. Describe it using both the visual frames and the speech content.\\
Visual frames: [Multiple screenshots are inserted here.]\\
Speech content: [Multiple ASR transcripts are inserted here.]\\
Note that the person shown in the stream may not be the host; another person may be broadcasting.\\
Describe the events and content of the stream in detail, using the following fields:\\
Content: Detailed visual information, including visible text, the overall visual impression, and tone.\\
People: The appearance, clothing, and state of the host and any other people participating in the stream, when applicable.\\
Scene: The location, decorations, background audio, and other environmental details.\\
Event: What the host is doing, including but not limited to gaming, talent performances, or conversational interaction.\\
Atmosphere: The atmosphere of the stream, such as lively or calm.\\
Highlights: Whether the segment captures a climax of the stream, such as a dance, a PK session, or intense viewer interaction.\\
Target Users: Profiles of viewers who may be interested in the stream.\\
Describe each field in one paragraph. Return only the following JSON object:\\
\{\\
\hspace*{1.5em}"Content": "xxx",\\
\hspace*{1.5em}"People": "xxx",\\
\hspace*{1.5em}"Scene": "xxx",\\
\hspace*{1.5em}"Event": "xxx",\\
\hspace*{1.5em}"Atmosphere": "xxx",\\
\hspace*{1.5em}"Highlights": "xxx",\\
\hspace*{1.5em}"Target Users": "xxx"\\
\}\\
Ensure that the JSON can be parsed directly by Python or another programming language, with no missing, duplicated, or additional fields. Preserve sufficient visual details, attributes, and spatial relations, and keep the description consistent with basic common sense. Output only the JSON object, without a Markdown code block or any additional explanation.
\par
\endgroup
\end{minipage}}
\end{table*}

\section{Qualitative Comparisons}
\label{app:qualitative}

We compare five models on four live-stream examples using the same multimodal input and structured generation prompt.
Each table shows three sampled frames followed by complete English translations of the generated Event and Target Users fields.
Business addresses and personal details are omitted without changing the semantic content.
Green bold text marks recommendation-relevant details captured by RecoReward, whereas red bold text marks incorrect, unsupported, or overly generic details in the other outputs.
Uncolored text is provided for completeness and is not assigned a correctness judgment.
The examples were selected to illustrate differences in semantic emphasis and are not a random sample; the aggregate performance claims remain based on the matched-tower evaluation in Section~\ref{sec:rq1}.

\paragraph{Local food sales.}
RecoReward retains the concrete shop-directed action, the discussion of product size and fat content, and the informal interaction style.
The two commercial models also recover several useful transactional details, whereas the remaining open-source models confuse braised food with roasted or barbecued meat and produce broader target-user descriptions.

\paragraph{Decorative-painting sales.}
RecoReward jointly describes the visible pointing action, the pearl-finish craft, the feng shui layout, and the adjustment of a painting to different wall layouts.
GPT-5, Gemini-3.1-Pro, and Qwen3-VL-8B also retain much of the product narration, but some target-user descriptions extend to collectors, investors, or high-end consumers without direct evidence.
The InternVL3.5 output retains the topic but omits much of the craft and placement information.

\paragraph{Daily-life conversation.}
RecoReward preserves the distinction between teaching new learners and hiring a professional cook, while also representing the informal interaction style.
Qwen3-VL-8B incorrectly merges the two activities into teaching learners to cook, and several other descriptions map the stream to broad learning or household-interest groups.
GPT-5 and Gemini-3.1-Pro provide more accurate event summaries, but their target-user descriptions are less focused on the conversational style of the stream.

\paragraph{Sneaker sales.}
RecoReward follows the sales sequence in detail, including repositioning the shoes, comparing the black and white variants, checking inventory, and explaining the packaging and inspection policy.
The other models recover the main transaction but either omit this sequence or add demographic and product details that are not supported by the input.

\section{Recall Metric Definitions}
\label{app:experimental-details}

Let \(r_q\) denote the rank of the target live stream for query \(q\), and let \(\mathcal{Q}\) denote the set of evaluation queries.
We compute Hit Rate, Normalized Discounted Cumulative Gain, and Mean Reciprocal Rank as
\begin{align}
\operatorname{HR@K}
&=
\frac{1}{|\mathcal{Q}|}
\sum_{q\in\mathcal{Q}}\mathbb{I}[r_q\leq K],
\label{eq:hr}\\
\operatorname{NDCG@K}
&=
\frac{1}{|\mathcal{Q}|}
\sum_{q\in\mathcal{Q}}
\frac{\mathbb{I}[r_q\leq K]}{\log_2(r_q+1)},
\label{eq:ndcg}\\
\operatorname{MRR}
&=
\frac{1}{|\mathcal{Q}|}
\sum_{q\in\mathcal{Q}}\frac{1}{r_q}.
\label{eq:mrr}
\end{align}
We evaluate HR and NDCG at \(K\in\{10,64,128\}\).

\section{Limitations}
\label{app:limitations}

First, RAS is constructed from observational user behavior.
Historical target users approximate future users, whereas non-target users are active users without observed positive interactions rather than exposed negatives.
Exposure patterns, item popularity, and prior recommender policies may therefore affect both groups, so the resulting score should not be interpreted as causal user preference.

Second, RAS evaluates user-selective compatibility in a frozen two-tower space rather than factual accuracy or general description quality.
A high score can favor details that separate user groups without guaranteeing that every detail is fully supported by the multimodal input, and stronger RL optimization may amplify patterns specific to the proxy reward.
Independent factuality and grounding evaluation is needed to determine how this trade-off changes with optimization strength.

Third, the empirical study is limited to live-stream recommendation on one platform, a fixed structured prompt, and matched two-tower recall evaluation.
The three reported seeds vary evaluator initialization rather than policy training, and each ablation setting uses one actor run.
The results therefore do not establish generalization to other recommendation domains, languages, output schemas, or recommender architectures, and the ablation explanations should be interpreted descriptively.
Future work should include matched actor replications and evaluation with alternative recall and ranking models.

\begin{table*}[p]
\centering
\begingroup
\setlength{\abovecaptionskip}{0pt}
\setlength{\belowcaptionskip}{3pt}
\noindent
\begin{minipage}{\textwidth}
  \centering
  \captionof{table}{Complete Event and Target Users outputs for a local smoked-meat stall.}
  \label{tab:case-smoked-meat}
  \scriptsize
  \setlength{\tabcolsep}{2pt}
  \renewcommand{\arraystretch}{0.90}
  \begin{tabular}{@{}p{0.11\linewidth}p{0.50\linewidth}p{0.31\linewidth}@{}}
    \multicolumn{3}{c}{%
      \includegraphics[height=0.92in]{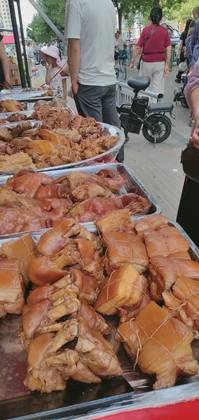}\hspace{0.6em}
      \includegraphics[height=0.92in]{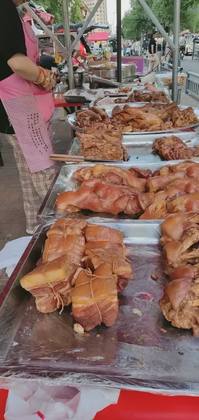}\hspace{0.6em}
      \includegraphics[height=0.92in]{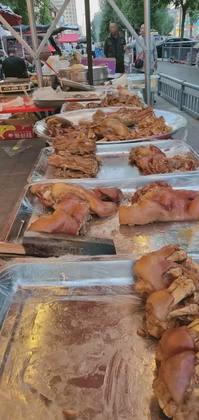}} \\
    \toprule
    Model & Event & Target Users \\
    \midrule
    GPT-5
    & The stream presents an on-site sales demonstration at a braised- and smoked-meat stall. The host or a clerk states that the food is prepared and replenished on site, is available immediately, and contains no added coloring or additives. The description provides the shop name and location, explains meat-selection criteria, weighs and packs orders for visiting customers, interacts with viewers, recommends several meat and side-dish combinations, and discusses taste and portion size.
    & Local residents and market visitors who enjoy braised meat, smoked meat, and other cooked foods; consumers who prefer generous portions, rich flavors, and home-style food; households that value fresh preparation without additives; nearby users interested in pickup or an in-store tasting; and \qualbad{restaurant operators or households seeking large cooked-food platters for gatherings}. \\
    \addlinespace[1.5pt]
    Gemini-3.1-Pro
    & The host uses a first-person view to show the cooked foods at the stall. The narration states that the smoked meat and bone-in products are cooked, prepared, and processed on site without coloring or additives, gives the specific shop location, and invites viewers to visit. The host also discusses the fat content and quality of the meat with customers or viewers.
    & Food enthusiasts, especially users who enjoy traditional cooked foods such as braised and smoked meat, together with residents near the stall who \qualbad{enjoy browsing street markets and searching broadly for local snacks}. \\
    \addlinespace[1.5pt]
    Qwen3-VL-8B
    & The stallholder cuts and sells \qualbad{braised pig trotters and pig-head meat} while customers choose and purchase products. The stallholder cuts meat at the counter, and several trays are empty or nearly empty, showing an active sales process.
    & Users interested in street food and braised dishes, especially young and middle-aged users who enjoy pig trotters and pig-head meat, as well as \qualbad{tourists or residents seeking a general street-market experience}. \\
    \addlinespace[1.5pt]
    InternVL3.5-8B
    & The stallholders display and sell \qualbad{several roasted-pork products}, describe their characteristics and freshness, and process customer orders.
    & \qualbad{General viewers interested in roasted meat and night-market or outdoor-market content}. \\
    \addlinespace[1.5pt]
    \rowcolor{QualitativeRowBlue}
    RecoReward-4B
    & The host displays and sells braised meat at an outdoor stall, including pig trotters, pork belly, and whole chicken. She cuts the products while discussing their source and price, responds to jokes about body weight and product size, \qualgood{directs viewers to the specific shop}, interacts with other stallholders, helps pack purchases, and \qualgood{continues the discussion of product size, fat content, and viewer comments}.
    & Middle-aged and lower-tier-market users interested in \qualgood{regional food and unpolished daily-life content}, together with general users who prefer \qualgood{authentic, minimally edited, conversational live streams}. \\
    \bottomrule
\end{tabular}
\end{minipage}
\par\vspace{3pt}

\noindent
\begin{minipage}{\textwidth}
  \centering
  \captionof{table}{Complete Event and Target Users outputs for a landscape-painting sales stream.}
  \label{tab:case-landscape-painting}
  \scriptsize
  \setlength{\tabcolsep}{2pt}
  \renewcommand{\arraystretch}{0.90}
  \begin{tabular}{@{}p{0.11\linewidth}p{0.50\linewidth}p{0.31\linewidth}@{}}
    \multicolumn{3}{c}{%
      \includegraphics[height=0.92in]{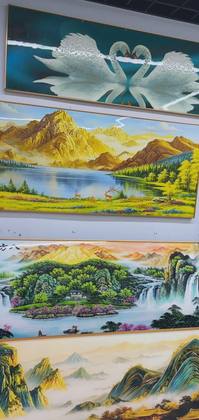}\hspace{0.6em}
      \includegraphics[height=0.92in]{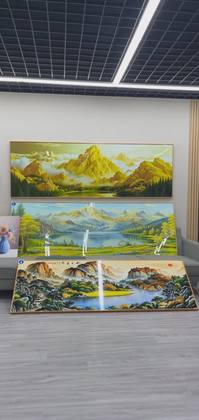}\hspace{0.6em}
      \includegraphics[height=0.92in]{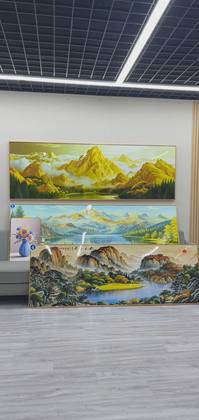}} \\
    \toprule
    Model & Event & Target Users \\
    \midrule
    GPT-5
    & The host or a staff member promotes the craft and meaning of traditional paintings, focusing on a work titled ``Golden Mountains.'' The description states that the painting suits different wall surfaces and has no placement taboo, reports strong offline sales, explains the pearl-finish proofing process and the cost of repeated revisions, and interprets the feng shui layout of the Azure Dragon, White Tiger, Vermilion Bird, Black Tortoise, and central treasure basin. It also lists suitable settings, including living rooms, sofa walls, television walls, studies, tea rooms, and offices, and explains the associated meanings of prosperity, wealth, and blessing.
    & Household users who value Chinese aesthetics and feng shui, homeowners decorating living rooms or feature walls, users furnishing tea rooms or studies, buyers for offices and reception spaces, consumers sensitive to craft texture and reflective finishes, \qualbad{art collectors}, and young or middle-aged users and merchants who value auspicious meanings and spatial feng shui. \\
    \addlinespace[1.5pt]
    Gemini-3.1-Pro
    & The host conducts a sales stream for decorative paintings and promotes a work titled ``Golden Mountains.'' He explains suitable placements such as a central living-room wall or an office, reports more than 3,000 offline orders per day, states that proofing cost RMB 66,000 and required more than 20 revisions, and emphasizes the Azure Dragon, White Tiger, Vermilion Bird, Black Tortoise, and treasure-basin symbolism to encourage purchase.
    & Users seeking home decoration, especially \qualbad{middle-aged users selected mainly from the visual style}, users who favor brightly colored landscape paintings with a diamond-like finish, and business owners who need office or tea-room decoration and value feng shui. \\
    \addlinespace[1.5pt]
    Qwen3-VL-8B
    & The host introduces several landscape paintings through narration and focuses on ``Golden Mountains.'' The description covers the Azure Dragon, White Tiger, Vermilion Bird, and Black Tortoise symbolism for prosperity, reports more than 3,000 offline orders per day, describes a pearl-finish proofing cost of RMB 66,000, and lists suitable placements such as living rooms, studies, and offices.
    & Users interested in home decoration, feng shui culture, and art collecting, particularly \qualbad{high-end consumers and collectors} seeking paintings for living rooms, studies, or offices, and users who value feng shui layouts and auspicious meanings. \\
    \addlinespace[1.5pt]
    InternVL3.5-8B
    & The host introduces and promotes the landscape paintings, explains their meanings and characteristics, and states that they are popular products both online and offline.
    & Users interested in home decoration and art collecting, especially viewers who \qualbad{primarily enjoy landscape paintings and natural scenery}. \\
    \addlinespace[1.5pt]
    \rowcolor{QualitativeRowBlue}
    RecoReward-4B
    & In a room used as both a living room and a study, the host presents several large Chinese landscape paintings mounted on the wall or placed on the floor. He \qualgood{uses a green pointer to explain their composition and pearl-finish craft}, discusses prices and revision costs, compares works such as ``Golden Mountains'' and a red-maple landscape, and \qualgood{explains the Azure Dragon, White Tiger, Vermilion Bird, and Black Tortoise layout}. The host also shows numbered details and close-up views, responds to questions about price and matching, and \qualgood{demonstrates how position and viewing angle can be adjusted for different wall layouts}.
    & Middle-aged and older users interested in \qualgood{traditional culture, feng shui meanings, and home decoration}; users seeking affordable handicrafts or folk-style decoration; and viewers interested in informal, conversational sales streams. \\
    \bottomrule
\end{tabular}
\end{minipage}
\endgroup
\end{table*}

\begin{table*}[!t]
\centering
\begingroup
\setlength{\abovecaptionskip}{0pt}
\setlength{\belowcaptionskip}{3pt}
\noindent
\begin{minipage}{\textwidth}
  \centering
  \captionof{table}{Complete Event and Target Users outputs for a daily-life chat stream.}
  \label{tab:case-daily-chat}
  \scriptsize
  \setlength{\tabcolsep}{2pt}
  \renewcommand{\arraystretch}{0.94}
  \begin{tabular}{@{}p{0.11\linewidth}p{0.50\linewidth}p{0.31\linewidth}@{}}
    \multicolumn{3}{c}{%
      \includegraphics[height=0.92in]{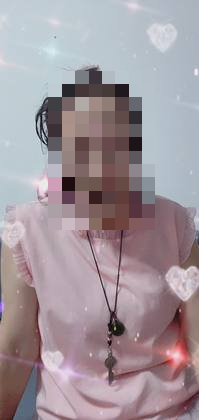}\hspace{0.6em}
      \includegraphics[height=0.92in]{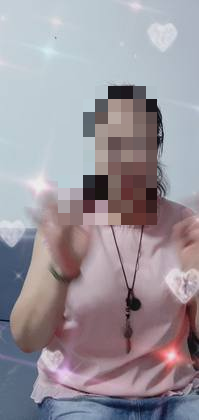}\hspace{0.6em}
      \includegraphics[height=0.92in]{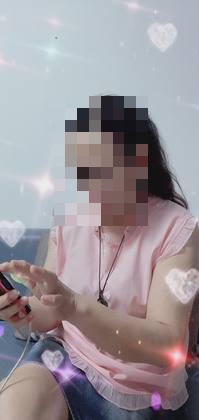}} \\
    \toprule
    Model & Event & Target Users \\
    \midrule
    GPT-5
    & The stream consists mainly of casual interaction. The host says that she will end the stream, take a shower, and return the next day. She explains that new friends or learners have arrived and that teaching them can delay her arrival in the stream, promises to make up for missed time, and says that she hired a professional cook while she buys the ingredients and organizes the learners and household tasks. She counts points on her fingers, claps for emphasis, checks her phone, and gives regular viewers a status update rather than presenting products or games.
    & Users who enjoy companion-style chat; young and middle-aged women interested in daily updates and emotional interaction; \qualbad{users interested in a learning community or offline instruction}; and regular viewers who prefer daily-life content and a long-term connection with the host. \\
    \addlinespace[1.5pt]
    Gemini-3.1-Pro
    & The host chats with viewers, says that she has remained at home and will take a shower after the stream, and reassures viewers not to be upset. She explains that new learners can delay her replies or limit the attention she gives to the stream, says that a professional was hired to cook while she teaches, and announces that the learner will graduate on the fourth day.
    & Fans interested in the daily life of the host, users following \qualbad{the training or learning content as a general topic}, and viewers who enjoy authentic daily records and casual conversation. \\
    \addlinespace[1.5pt]
    Qwen3-VL-8B
    & The woman explains that she is waiting at home and will take a shower after the stream. She previews the schedule for new learners, \qualbad{states that she teaches the new learners to cook}, thanks viewers for their support, and says that the learner will graduate on the fourth day.
    & Young women interested in daily-life sharing, emotional interaction, and learning content, particularly users interested in \qualbad{cooking and learning practical life skills}. \\
    \addlinespace[1.5pt]
    InternVL3.5-8B
    & The woman states that she is waiting at home and plans to take a shower. She also says that new friends or learners have arrived, that she needs to guide their study, and that a professional was hired to cook.
    & \qualbad{Young women interested in daily-life sharing, cooking, and learning-sharing content}, together with viewers who prefer warm and friendly live streams. \\
    \addlinespace[1.5pt]
    \rowcolor{QualitativeRowBlue}
    RecoReward-4B
    & Sitting on a sofa at home, the host chats with viewers about visiting friends and new learners. She checks and operates a red phone, gestures to emphasize her points, adjusts her hair, and rests her chin on one hand. She also responds to viewers who ask whether she is upset or will make up for missed time, and \qualgood{separately describes the professional cooking arrangement and the progress toward learner graduation} while continuing the everyday conversation.
    & Middle-aged users interested in \qualgood{authentic, informal daily-life conversation}, together with users who prefer non-professional entertainment and \qualgood{unpolished, conversational interaction}. \\
    \bottomrule
  \end{tabular}
\end{minipage}
\par\vspace{10pt}

\noindent
\begin{minipage}{\textwidth}
  \centering
  \captionof{table}{Complete Event and Target Users outputs for a sneaker sales stream.}
  \label{tab:case-sneaker-sales}
  \scriptsize
  \setlength{\tabcolsep}{2pt}
  \renewcommand{\arraystretch}{0.94}
  \begin{tabular}{@{}p{0.11\linewidth}p{0.50\linewidth}p{0.31\linewidth}@{}}
    \multicolumn{3}{c}{%
      \includegraphics[height=0.92in]{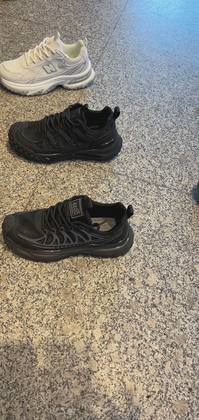}\hspace{0.6em}
      \includegraphics[height=0.92in]{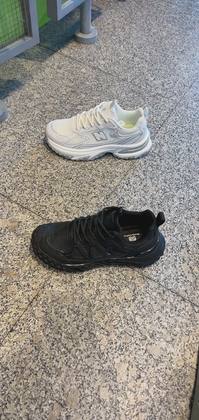}\hspace{0.6em}
      \includegraphics[height=0.92in]{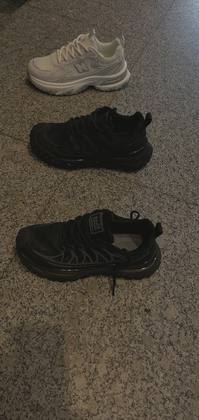}} \\
    \toprule
    Model & Event & Target Users \\
    \midrule
    GPT-5
    & The host sells New Balance-style sneakers and explains their sizes, inventory, warranty, breathable mesh, summer suitability, and standard fit. She reports that the front pair remains only in size 41, while the rear black pairs and the middle white pair remain in several adjacent sizes. She also promotes a price reduction from RMB 159 to RMB 78 and directs viewers to the flash-purchase link.
    & Price-conscious sports and casual-shoe consumers; commuters and walkers seeking breathable summer footwear; live-commerce buyers who value warranties and in-store verification; and \qualbad{young men and women selected mainly from the visual style} who wear sizes 41--44 and prefer either black outdoor-style or white chunky sneakers. \\
    \addlinespace[1.5pt]
    Gemini-3.1-Pro
    & The host sells New Balance sneakers, describing their white and black color variants, sizes, inventory, original labels, boxes, and tags, eligibility for in-store verification, warranty policy, and promotional price.
    & Consumers interested in sneakers, particularly those who prefer retro chunky designs and value brand style, competitive prices, quality assurance, and discounts, together with \qualbad{a broad male-and-female demographic}. \\
    \addlinespace[1.5pt]
    Qwen3-VL-8B
    & The host sells New Balance sneakers and describes their sizes, colors, \qualbad{whether they are new or used}, original labels, boxes, and tags, six-month warranty, original price of RMB 159, promotional price of RMB 78, and standard fit.
    & Consumers seeking sneakers, especially \qualbad{young users inferred without supporting evidence}, together with buyers who value brand, quality, standard sizing, and discounted prices. \\
    \addlinespace[1.5pt]
    InternVL3.5-8B
    & The host introduces and displays New Balance sneakers, emphasizing standard sizing, brand quality, and the promotional price.
    & \qualbad{General consumers interested in sneakers, especially viewers who follow the New Balance brand}. \\
    \addlinespace[1.5pt]
    \rowcolor{QualitativeRowBlue}
    RecoReward-4B
    & The host places several New Balance sneakers on the floor and shows their labels, mesh material, and sole design. She \qualgood{repositions the shoes to compare the black and white variants}, checks the remaining inventory, responds to questions about size and authenticity, and \qualgood{explains that the products include the original labels and boxes and can be inspected in a retail store}. She then discusses the promotional price and encourages viewers to purchase the remaining stock.
    & \qualgood{Price-sensitive consumers seeking affordable sneakers}, including young users interested in affordable domestic-trend or brand-alternative styles, students and budget-conscious young men seeking brand appeal, and niche viewers interested in bargain or second-hand shopping.
    \\
    \bottomrule
  \end{tabular}
\end{minipage}
\endgroup
\end{table*}

\end{document}